\documentclass[11pt,a4paper]{article}
\pdfoutput=1

\usepackage{comment}
\usepackage{jheppub}

\usepackage{physics}

\usepackage{subfig}

\newcommand{\be}{\begin{equation}}
\newcommand{\ee}{\end{equation}}
\newcommand{\bea}{\begin{eqnarray}}
\newcommand{\eea}{\end{eqnarray}}
\newcommand{\nn}{\nonumber}
\newcommand{\pd}{\partial}
\newcommand{\MP}{M_\text{P}}
\newcommand{\zetamax}{\zeta_\text{max}}

\title{\centering Slow-roll inflation in Palatini $F(R)$ gravity}

\author[a,b]{Christian Dioguardi}
\author[b]{Antonio Racioppi}
\author[b]{Eemeli Tomberg}

\affiliation[a]{Tallinn University of Technology, Akadeemia tee 23, 12618 Tallinn, Estonia}
\affiliation[b]{National Institute of Chemical Physics and Biophysics, R\"avala 10, 10143 Tallinn, Estonia}

\emailAdd{christian.dioguardi@kbfi.ee}
\emailAdd{antonio.racioppi@kbfi.ee}
\emailAdd{eemeli.tomberg@kbfi.ee }

\abstract{We study single field slow-roll  inflation in the presence of $F(R)$ gravity in the Palatini formulation. In contrast to metric $F(R)$, when rewritten in terms of an auxiliary field and moved to the Einstein frame, Palatini $F(R)$ does not develop a new dynamical degree of freedom. However, it is not possible to solve analytically the constraint equation of the auxiliary field for a general $F(R)$. We propose a method that allows us to circumvent this issue and compute the inflationary observables. We apply this method to test scenarios of the form $F(R) = R + \alpha R^n$ and find that, as in the previously known $n=2$ case, a large $\alpha$ suppresses the tensor-to-scalar ratio $r$. 
We also find that models with $F(R)$ increasing faster than $R^2$ for large $R$ suffer from numerous problems. 

}


\begin{document}
\maketitle

\section{Introduction} \label{sec:Introdution}

Past and recent observations of the cosmic microwave background radiation (CMB) support the flatness and homogeneity of the Universe  at large scales. Such properties can be explained by assuming an accelerated expansion during the very early Universe \cite{Starobinsky:1980te,Guth:1980zm,Linde:1981mu,Albrecht:1982wi}. This inflationary era is also able to generate and preserve the primordial inhomogeneities which generated the subsequent large-scale structure that we observe. In its minimal version, inflation is usually formulated by adding to the Einstein-Hilbert action one scalar field, the inflaton, whose energy density induces the near-exponential expansion. 

Recently, the BICEP/Keck Array experiment \cite{BICEP:2021xfz} has reduced even more the available parameter space, disfavoring most of the minimal inflationary models. On the other hand, the maybe most popular inflationary setup, the Starobinsky model~\cite{Starobinsky:1980te}, still lies in the allowed region. Such a model involves non-minimal gravity and it can be equivalently described by the addition of a $R^2$ term to the Einstein-Hilbert action or by a scalar field non-minimally coupled to gravity  (e.g. \cite{Jarv:2016sow} and references therein).

However, in the context of  non-minimally  coupled  theories there is more than one \emph{choice} of the dynamical degrees of freedom. 
In  the  common metric formulation, the  metric tensor is the only dynamical degree of freedom, while the connection is fixed to be the Levi-Civita one.  On the other hand,  in the Palatini formulation, both the metric and the connection are independent variables. Their corresponding equations of motion (EoMs) will dictate the eventual relation between the two variables. When the action is linear in the curvature scalar, the two formalisms  lead to equivalent theories (i.e. the Levi-Civita connection arises from the solution of one EoM), otherwise the theories are completely different \cite{Bauer:2008zj} and lead to different phenomenological predictions, as recently investigated in e.g. \cite{Koivisto:2005yc,Tamanini:2010uq,Bauer:2010jg,Rasanen:2017ivk,Tenkanen:2017jih,Racioppi:2017spw,Markkanen:2017tun,Jarv:2017azx,Racioppi:2018zoy,Kannike:2018zwn,Enckell:2018kkc,Enckell:2018hmo,Rasanen:2018ihz,Bostan:2019uvv,Bostan:2019wsd,Carrilho:2018ffi,Almeida:2018oid,Takahashi:2018brt,Tenkanen:2019jiq,Tenkanen:2019xzn,Tenkanen:2019wsd,Kozak:2018vlp,Antoniadis:2018yfq,Antoniadis:2018ywb,Gialamas:2019nly,Racioppi:2019jsp,Rubio:2019ypq,Lloyd-Stubbs:2020pvx,Das:2020kff,McDonald:2020lpz,Shaposhnikov:2020fdv,Enckell:2020lvn,Jarv:2020qqm,Gialamas:2020snr,Karam:2020rpa,Gialamas:2020vto,Karam:2021wzz,Karam:2021sno,Gialamas:2021enw,Annala:2021zdt,Racioppi:2021ynx,Cheong:2021kyc,Mikura:2021clt,Ito:2021ssc,Racioppi:2021jai}.

In particular, there is a dramatic difference between the metric and the Palatini formulations when an $R^2$ term is added to a single scalar field inflationary action. In the metric case, we obtain a bi-field inflationary setup (e.g. \cite{Kannike:2015apa,Karam:2018mft} and references therein), while in the Palatini case we still obtain a single field scenario \cite{Enckell:2018hmo}. In the latter case, it is remarkable that the presence of the $R^2$ term leaves essentially unchanged all the phenomenological parameters except for the tensor-to-scalar ratio $r$, which can be arbitrarily lowered by increasing the coupling in front of the $R^2$ term. Motivated by the results of \cite{Enckell:2018hmo}, one wonders if any other $F(R)$ in the Palatini formulation can produce analogous results. However, such models are more complicated to study than $F(R)=R+\alpha R^2$. When studying $F(R)$ theories, it is common to use a representation via an auxiliary field and move the problem to the Einstein frame. In Palatini $F(R)=R+\alpha R^2$ the EoM for the auxiliary field is independent of $\alpha$ and quite simple to solve \cite{Enckell:2018hmo}. Unfortunately, the same does not happen for any generic $F(R)$, where the EoM for the auxiliary field may not be analytically solvable at all. The purpose of our work is to present a method that allows the computation of inflationary predictions even in such a case.

The article is organized as follows. In section \ref{sec:SR}, we introduce the theory of a single field inflaton in the presence of Palatini $F(R)$ gravity and develop a new method that allows slow-roll computations even when it is not possible to solve exactly the EoM of the auxiliary field. In section \ref{sec:test}, we apply our method to test scenarios of the form $F(R) = R + \alpha R^n$ with arbitrary $n>1$. Finally, in section \ref{sec:beyond_SR}, we check the behaviour of the aforementioned scenarios beyond the slow-roll approximation, in particular in the high $R$ limit. We present our conclusions in section \ref{sec:Conclusions}. In appendix~\ref{sec:comparison_to_other_RN_computation}, we compare our results to an earlier work \cite{Bekov:2020dww} that discussed similar models.

\section{Slow-roll computations} \label{sec:SR} 

We start by considering the following action for a real scalar inflaton $\phi$ minimally coupled to a $F(R)$ gravity (in Planck units: $\MP=1$):
\be \label{eq:actionFR}
  S_J = \int \dd^4 x \sqrt{-g_J} \left[\frac{1}{2} F(R(\Gamma)) - \frac{1}{2}  k(\phi) g_J^{\mu\nu} \partial_\mu \phi \partial_\nu \phi - V(\phi) \right] \,,
\ee
where $k(\phi)>0$ is the non-minimal kinetic function for the inflaton and $V(\phi)$ its positive scalar potential\footnote{The study applies also to theories with $\phi$ non-minimally coupled to gravity, if it is possible to perform a frame transformation and cast the action in the form of \eqref{eq:actionFR} (or equivalently \eqref{eq:action:zeta:J} e.g. \cite{Jarv:2016sow,Jarv:2020qqm}). The cases where this is not possible need to be investigated separately. We postpone such a study to a future work.}. We stress that we are considering the Palatini formulation of gravity by using the notation $R(\Gamma)$, where $R$ is the curvature scalar and $\Gamma^\rho_{\mu\nu}$ is the connection in the Jordan frame.
As is customary, we rewrite the $F(R)$ term using the auxiliary field $\zeta$
\be \label{eq:action:zeta:J}
  S_J = \int \dd^4 x \sqrt{-g}_J \left[\frac{1}{2} \left(F(\zeta)+F'(\zeta) \left(R(\Gamma)-\zeta \right) \right) - \frac{1}{2}  k(\phi)  g_J^{\mu\nu} \partial_\mu \phi \partial_\nu \phi - V(\phi) \right] \, ,
\ee
where $F'(\zeta)=\partial F/\partial \zeta$. Then, we move the theory to the Einstein frame via the Weyl transformation
\be \label{eq:Weyl}
  g_{E\, \mu \nu} = F'(\zeta) \ g_{J \, \mu \nu} \, ,  
\ee
which leads to the action
\be \label{eq:action:zeta:E}
  S_E = \int \dd^4 x \sqrt{-g_E} \left[ \frac{\MP^2}{2} R_E - \frac{1}{2} g_E^{\mu\nu} \partial_\mu \chi \partial_\nu \chi - U(\chi,\zeta) \right] \, ,
\ee
where the canonically normalized scalar $\chi$ is defined by
\be \label{eq:dchidphi}
 \frac{\pd \chi}{\pd \phi} = \sqrt{\frac{k(\phi)}{F'(\zeta)}} \, ,
\ee
and the full scalar potential is
\be \label{eq:Uchizeta}
  U(\chi,\zeta) = \frac{V(\phi(\chi))}{F'(\zeta)^2} - \frac{F(\zeta)}{2 F'(\zeta)^2} + \frac{\zeta}{2 F'(\zeta)} \, .
\ee
By varying \eqref{eq:action:zeta:E} with respect to $\zeta$, we get its EoM in the Einstein frame,
\be
2 F(\zeta ) -\zeta  F'(\zeta ) -k(\phi ) \, \partial^\mu \phi \partial_\mu \phi \, F'(\zeta ) -4 V(\phi ) = 0 \, , \label{eq:EoMzetafull}
\ee
where we assumed $F'(R),F''(R) \neq 0$. The standard procedure would be now to solve \eqref{eq:EoMzetafull}, determine the solution for the auxiliary field as $\zeta(\phi, \partial^\mu \phi \partial_\mu \phi)$ and insert it back into the action \eqref{eq:action:zeta:E}. However, for a generic $F(R)$, we cannot expect \eqref{eq:EoMzetafull} to be explicitly solvable, even though it should still be satisfied. On the other hand, we might still be able to perform inflationary computations in the slow-roll approximation.
Assuming that slow-roll conditions are satisfied (i.e. $ g_J^{\mu\nu} \partial_\mu \phi \partial_\nu \phi \ll V(\phi)$)\footnote{We will go back to the full evolution of the system in Section \ref{sec:beyond_SR} and discuss the slow-roll conditions in more detail.}, we can write the EoM  \eqref{eq:EoMzetafull} as
\be \label{eq:EoMzeta}
 G(\zeta) =  V(\phi) \, ,
\ee
with 
\be \label{eq:G}
G(\zeta) \equiv \frac{1}{4} \left[ 2 F(\zeta) - \zeta F'(\zeta) \right] \, .
\ee
In the slow-roll approximation we still cannot always provide explicit solutions for $\zeta$ for a given $F(R)$.
However, it is possible to perform inflationary computations by using the auxiliary field as a computational variable and using \eqref{eq:EoMzeta} as a constraint. First of all, by using  \eqref{eq:EoMzeta}, we replace $V(\phi)$ with $G(\zeta)$ in \eqref{eq:Uchizeta}, obtaining a scalar potential\footnote{Note that $U(\zeta)$ is an actual scalar potential only when the kinetic term of $\phi$ is negligible like in slow-roll.} that depends only on the  auxiliary field $\zeta$:
\be \label{eq:Uzeta}
  U(\chi,\zeta) = \frac{1}{4} \frac{ 2 F(\zeta) - \zeta F'(\zeta) }{F'(\zeta)^2} - \frac{F(\zeta)}{2 F'(\zeta)^2} + \frac{\zeta}{2 F'(\zeta)} = \frac{1}{4} \frac{\zeta}{F'(\zeta)} \equiv U(\zeta) \, .
\ee
We stress that the last result implies that $\zeta>0$ in the slow-roll regime, since both the potential $U(\zeta)$ and $F'(\zeta)$ (controlling the signs of the Weyl transformation and the kinetic term) should be positive there.
It was shown in \cite{Enckell:2018hmo} that $F(R) = R + \alpha R^2$ gives an asymptotically flat potential, regardless of the initial $V(\phi)$. We can easily deduce from \eqref{eq:Uzeta} that no other asymptotic form of $F(R)$ than $F(R)\sim R^2$ can give such a result.

Let us proceed and perform inflationary computations. We need to compute the first derivative of $U(\zeta)$ with respect to $\chi$, the canonically normalized scalar field in the Einstein frame:
\bea \label{eq:dUdchi}
  \frac{\pd}{\pd \chi} U(\zeta) 
  &=& \frac{\pd \phi}{\pd \chi}\frac{\pd \zeta}{\pd \phi} \frac{\pd}{\pd \zeta} U(\zeta) = 
  \sqrt{\frac{F'(\zeta)}{k( V^{-1}(G) )}} \left(\frac{\pd G}{\pd \zeta} \frac{\pd V^{-1}}{\pd G}\right)^{\!-1} \frac{\pd U}{\pd \zeta} \, ,
 \label{eq:Vp}
\eea
where we used the chain rule for the derivative of composite functions together with \eqref{eq:dchidphi}, and $V^{-1}(G)$ is the inverse function of $V(\phi)$ defined via \eqref{eq:EoMzeta}. In the end, after using \eqref{eq:G} for $G$, we have a function of $\zeta$ only. Similarly, for a general function $f(\zeta)$, we have:
\begin{equation} \label{eq:general_zeta_derivatives}
    \frac{\partial}{\partial \chi} f(\zeta) = g(\zeta) \frac{\partial f(\zeta)}{\partial \zeta} \, ,
\end{equation}
where
\begin{equation} \label{eq:gzeta}
    g(\zeta) \equiv \frac{\partial \zeta}{\partial \chi}
    = \sqrt{\frac{F'(\zeta)}{k( V^{-1}(G) )}} \left(\frac{\pd G}{\pd \zeta} \frac{\pd V^{-1}}{\pd G}\right)^{\!-1} \, .
\end{equation}
 This derivative can be explicitly computed as long as $V$ is invertible. This allows us to easily express higher order derivatives:
\begin{equation} \label{eq:Udotdot}
    \frac{\partial^2}{\partial \chi^2} U(\zeta) = g(\zeta) \frac{\partial}{\partial \zeta} \qty(g(\zeta) \frac{\partial U}{\partial \zeta}) = gg'U' + g^2U'' \ , \ \dots
\end{equation}
From \eqref{eq:dUdchi} and \eqref{eq:Udotdot} we can derive the slow-roll parameters straightforwardly:
\be \label{eq:epsilon_zeta}
\epsilon(\zeta) = \frac{1}{2} \qty(\frac{\pd U/\pd\chi}{U})^2 = \frac{1}{2}g^2 \qty(\frac{U'}{U})^2 \, ,
\ee

\be \label{eq:eta_zeta}
\eta(\zeta) = \frac{\pd^2 U/\pd\chi^2}{U} =  \frac{gg'U'+ g^2 U''}{U} \, .
\ee
Hence, the equations for the CMB observables $r$, $n_s$, $A_s$ read:
\begin{eqnarray}
\label{eq:r_zeta}
r(\zeta) &=& 16 \epsilon(\zeta) = 8 g^2 \qty(\frac{U'}{U})^2 \, , \\
\label{eq:ns_zeta}
n_s(\zeta) &=& 1 + 2 \eta(\zeta) - 6 \epsilon(\zeta) = 1 + \frac{2g}{U^2} \qty(g'U'U + g U''U - 24 g U'^2) \, , \\
\label{eq:As_zeta}
A_s (\zeta) &=& \frac{U}{24\pi^2 \epsilon(\zeta)} = \frac{U^3}{12 \pi^2 g^2 U'^2} \, ,
\end{eqnarray}
where $A_s$ has to satisfy \cite{Planck2018:inflation}
\be
\label{eq:As:constraint}
\ln \left(10^{10} A_s \right) = 3.044 \pm 0.014   \, 
\ee
at the CMB scale.
Analogously, the number of e-folds becomes
\begin{equation} \label{eq:N_zeta}
    N_e = \int_{\chi_f}^{\chi_N} \frac{U}{\partial U / \partial \chi} \dd \chi = \int_{\zeta_f}^{\zeta_N}  \frac{U}{g^2 \, \partial U / \partial \zeta} \dd \zeta \, ,
\end{equation}
where value at the end of inflation, $\zeta_f$, is determined  by\footnote{ The careful reader might notice that $|\eta|=1$ could trigger the end of slow-roll before the actual end of the inflationary phase. However, this is never the case in our example scenarios below, at least for the parameter space that we considered.} $\epsilon(\zeta_f) = 1$. Equation \eqref{eq:N_zeta} determines the auxiliary field value $\zeta_N$ at the time a given scale leaves the horizon, corresponding to $N_e$. In our examples below, we take $N_e  \in [50,60]$ at CMB.

Before proceeding to examples, we will briefly comment on the mandatory requirements of our procedure in the following subsection.

\subsection{Requirements} \label{subsec:requirements}
First of all we need to satisfy the usual requirements of any non-minimal gravity model, i.e. reproducing Einstein gravity as a low energy limit and having the correct  positive sign in the Weyl transformation \eqref{eq:Weyl} (i.e. $F'(R)>0$). Then, we need to satisfy the constraint in eq. \eqref{eq:EoMzeta}. This induces several additional conditions. First,  $V(\phi)$ needs to be an invertible function, so that we can define the function $g(\zeta)$ in eq. \eqref{eq:gzeta}. The function $g(\zeta)$ must also be uniquely defined, therefore $G(\zeta)$ must be a bijective function, at least in a smaller domain that satisfies eq. \eqref{eq:EoMzeta}. Last but not least, since $V(\phi)$ is positive everywhere, the same must be true also for $G(\zeta)$ in the region of validity of the slow-roll approximation, i.e. $\zeta>0$. As we can see from eq. \eqref{eq:G}, this is not, in general, true for an arbitrary $F(R)$. The problem lies in the fact that $G(\zeta)$ is the difference of two generally positive terms $F(\zeta)$ and $\zeta F'(\zeta)$.
Assuming that $F(\zeta) \sim \zeta^n$ for very large positive $\zeta$, we can see from  \eqref{eq:G} that $G(\zeta)$ is positive in this limit only when $n \leq$ 2. If that is the case, since $G$ is continuous and behaves linearly around 0, $G$ will take all possible positive values ensuring the existence of a $\zeta$ that satisfies  \eqref{eq:EoMzeta} (and also \eqref{eq:EoMzetafull}). We can easily extend the same reasoning to functions $F$ that do not possess a monomial asymptotic behaviour, leading to the following summary of requirements for successful scenarios:
\begin{eqnarray}
 &&F(R) \sim R \quad \text{when } R \sim 0 \, ,  \label{eq:low:energy:limit}\\
 &&F'(R) > 0 \, ,  \label{eq:good:Weyl}\\
 &&V(\phi) \quad \text{invertible} \, , \label{eq:Vinvertible}\\
 &&G(\zeta) \quad \text{bijective} \, , \label{eq:Gbijective}\\
 &&G(\zeta)>0 \ \text{when} \ \zeta>0 \Rightarrow \ \lim_{R \to + \infty} \frac{F(R)}{R^2} \to \text{positive or null constant}  \, .  \label{eq:high:energy:limit}
\end{eqnarray}
While it is relatively easy to satisfy the first four constraints, the last condition reduces noticeably the number of available $F(R)$ models. However, when  \eqref{eq:high:energy:limit} is not satisfied, it is still possible to perform some inflationary computations if some additional constraints are realized. When \eqref{eq:high:energy:limit} does not hold but \eqref{eq:low:energy:limit} and \eqref{eq:good:Weyl} do, $G(\zeta)$ is a function bounded from above (in the real positive domain) with at least one local maximum. This means that \eqref{eq:EoMzeta} can be satisfied only for the $V(\phi)$ values that lie within the upper limit of $G(\zeta)$. In this case, inflationary computations can still make sense if slow-roll is realized within such a region and $G(\zeta)$ is treated as an effective description.

\begin{figure}[t]
    \centering
    \includegraphics[width=0.45\textwidth]{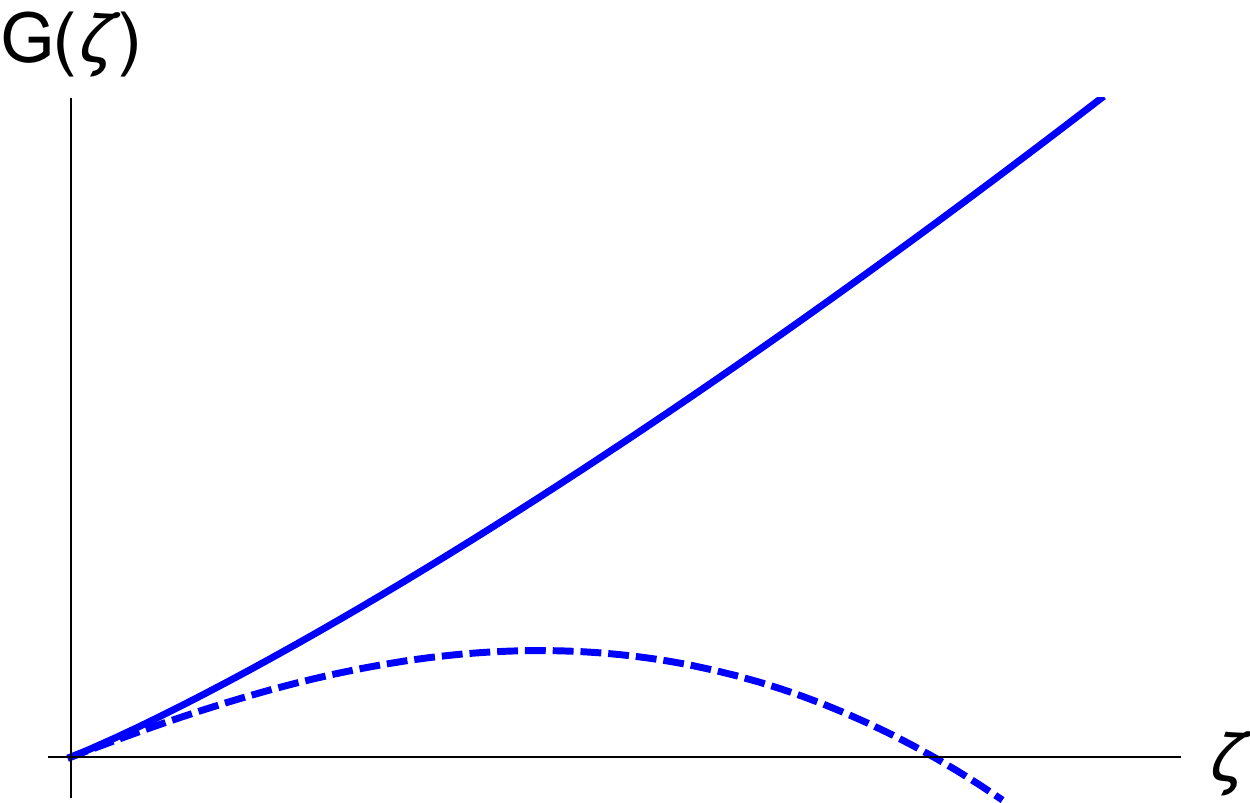}
    \qquad
    \includegraphics[width=0.45\textwidth]{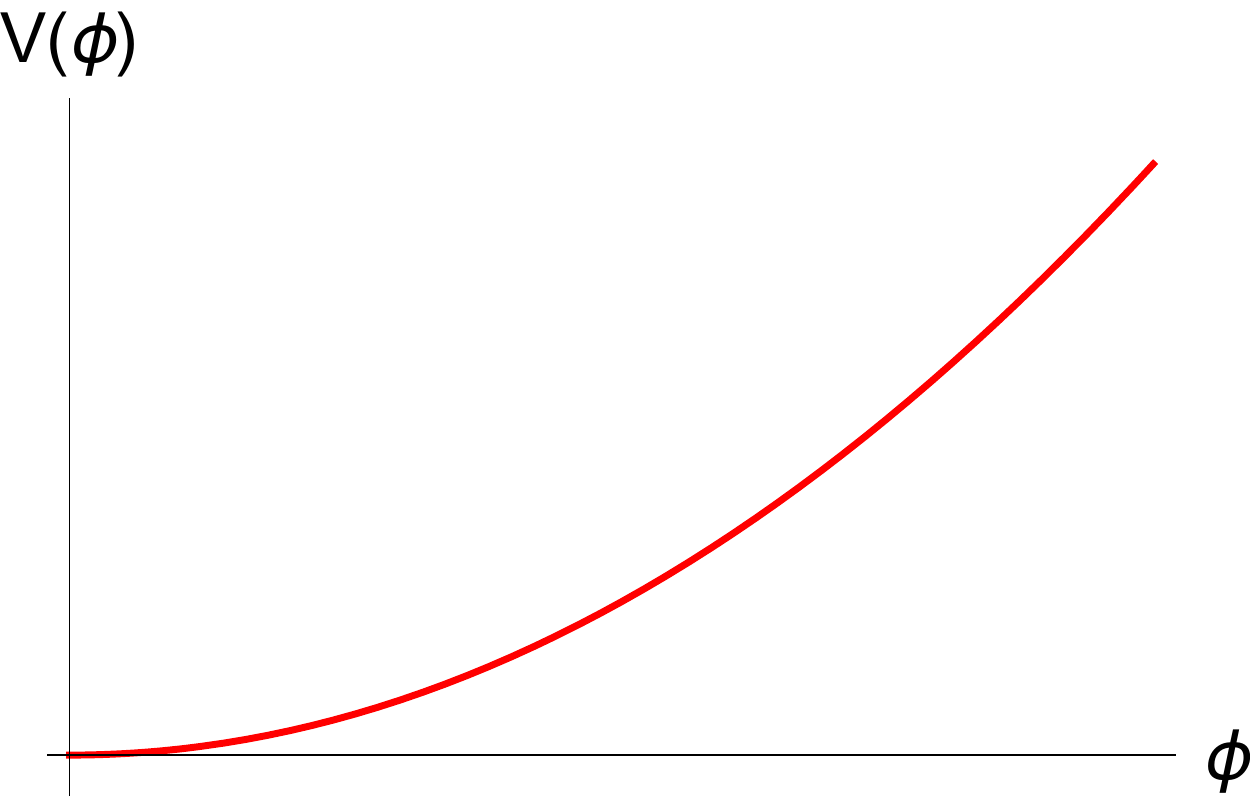}
    \caption{Reference plots of $G(\zeta)$ (left) and $V(\phi)=\phi^2$ (right) for $F(R)=R+R^n$ with $n=3/2$ (continuous) and $n=5/2$ (dashed). }%
    \label{fig:GvsV}
\end{figure}

In Fig. \ref{fig:GvsV}, we present a visual example of the issue. In the left panel we show a reference plot of $G(\zeta) $ for $F(R)=R+R^n$ with $n=3/2$ (continuous) and $n=5/2$ (dashed), while in the right panel we show a reference plot of $V(\phi)=\phi^2$. As we can see in the right panel, $V$ is covering all positive values in the $y$-axis as expected. The same happens in the left panel for $n=3/2$ i.e. when \eqref{eq:high:energy:limit} is satisfied.  Therefore, a $\zeta$ that satisfies \eqref{eq:EoMzeta} always exists. On the other hand, for $n=5/2$ i.e. when \eqref{eq:high:energy:limit} is not satisfied, $G$ presents a local maximum and then decreases towards negative values. Therefore, when the $V$ value is high enough, there is no real solution for $\zeta$ that can satisfy \eqref{eq:EoMzeta}. Hence, an effective description is the only available working option. The positivity and bijectivity of $G$ (together with condition \eqref{eq:low:energy:limit}) can be still realized within the origin and the local maximum of $G$. Therefore slow-roll must be realized within this interval in order to have at least a feasible inflationary model.

In the following section, we will present a numerical study considering test scenarios in both $n<2$ and $n>2$ configurations. After that, we will comment on the beyond slow-roll behaviour and see in more detail the problems that arise if \eqref{eq:high:energy:limit} is not satisfied for large $\zeta$.

\section{Test scenarios} \label{sec:test}
 
 In this section we test our method with 
 \be
   F(R)=R+\alpha R^n \, , \quad k(\phi)=1 \, , \quad V(\phi)= \frac{m^2}{2} \phi^2 \, . \label{eq:testscenario}
 \ee
 We consider two different scenarios\footnote{{In the limit $\alpha \to \infty$, our results should agree with those of \cite{Bekov:2020dww}, where the authors considered slow-roll in Palatini $F(R)$ models with $F(R)\sim R^n$. However, this does not happen because of a sign error in \cite{Bekov:2020dww}. We discuss the differences between our computation and theirs in appendix \ref{sec:comparison_to_other_RN_computation}.}}, $n<2$ and $n>2$. We stress that, during slow-roll, the positivity of both $U(\zeta)$  and $F'(\zeta)$ implies $\zeta>0$ and $\alpha>0$.
 Before proceeding, we also check that our procedure reproduces the results of \cite{Enckell:2018hmo} with $F(R)=R+\alpha R^2$ for any kind of $V(\phi)$. 

\subsection{$n=2$}
We can easily verify that for the $F(R) = R+\alpha R^2$ we get the same results as in \cite{Enckell:2018hmo}. These results can be cast in the following form:
 \begin{eqnarray}
  &&U = \frac{V}{1 + 8 \alpha V} = \frac{U^0}{1 + 8 \alpha U^0} \label{eq:U:R2} \, , \\
  &&r = \frac{r^0}{1 + 8 \alpha U^0} \, ,   \label{eq:r:R2} \\ 
  &&n_s = n_s^0 \, , \qquad 
  N_e = N_e^0 \, , \qquad
  A_s = A_s^0 \, , \label{eq:other:R2}
  \end{eqnarray} 
where $\dots^0$ means that the quantity is computed for $\alpha=0$. First of all, we check the scalar potential. Eq. \eqref{eq:G} becomes
\begin{eqnarray}
G(\zeta) &=& \frac{1}{4} \left[ 2 F(\zeta) - \zeta F'(\zeta) \right] = 
           \frac{1}{4} \left[ 2 \zeta + 2\alpha \zeta^2 - \zeta (1+2\alpha \zeta) \right]\nn\\
           &=& \frac{1}{4} \zeta = V(\phi)=U^0 \, ,
\end{eqnarray}
which leads to 
 \be 
  U = \frac{1}{4} \frac{\zeta}{F'(\zeta)}=\frac{\zeta }{4+8 \alpha  \zeta} = \frac{U^0}{1 + 8 \alpha U^0} \, ,
\ee
in agreement with eq. \eqref{eq:U:R2}. Let us check now the inflationary parameters, starting with $r$. First of all, we need to evaluate the $g$ function:
\be
 g = \sqrt{F'(\zeta)} \left(\frac{\pd G}{\pd \zeta} \frac{\pd V^{-1}}{\pd G}\right)^{\!-1} = \sqrt{1+2\alpha\zeta} \left(\frac{\pd V^{-1}}{\pd \zeta}\right)^{\!-1}  \, .
\ee
Therefore, the tensor-to-scalar ratio becomes
\bea
 r(\zeta) = 8 g^2 \qty(\frac{U'}{U})^2 = \left(\frac{\pd V^{-1}}{\pd \zeta}\right)^{\!-2} \frac{8} {\zeta^2} \left(\frac{1}{1+2 \alpha\zeta}\right) = \frac{r^0}{1 + 8 \alpha U^0} \, ,
\eea
again in agreement with the result of \cite{Enckell:2018hmo} in eq. \eqref{eq:r:R2}.
Analogously we can prove the validity of the remaining results in eq. \eqref{eq:other:R2}.

\subsection{$n<2$}

\begin{figure}[t]%
    \centering
    \subfloat[]{\includegraphics[width=0.45\textwidth]{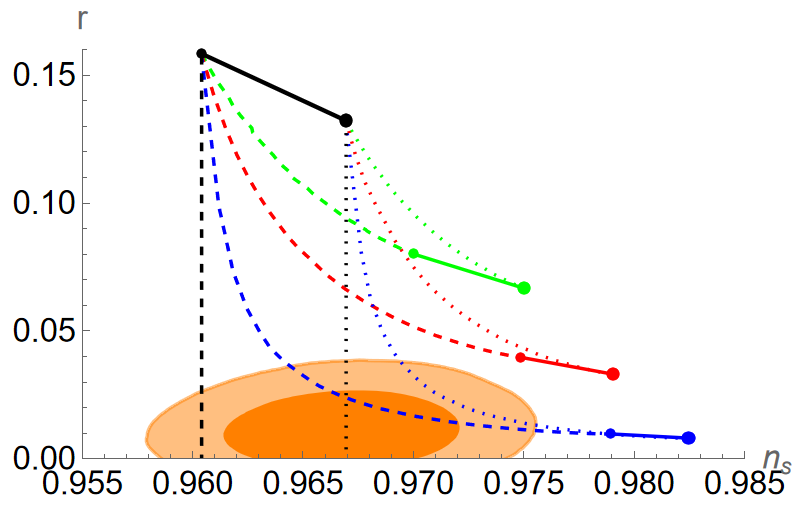}}%
    \qquad
    \subfloat[]{\includegraphics[width=0.45\textwidth]{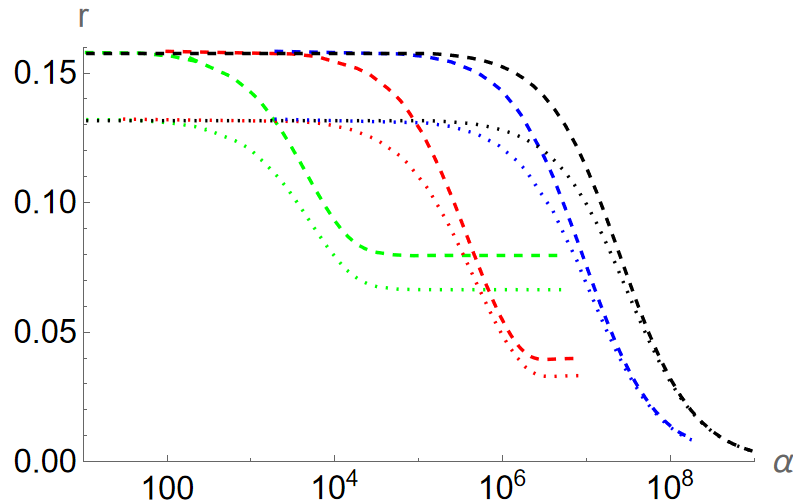}}%
    
    \subfloat[]{\includegraphics[width=0.45\textwidth]{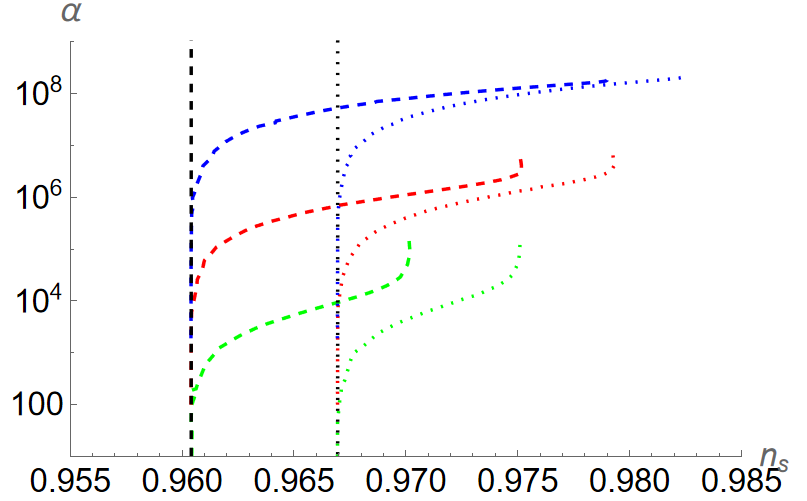}}%
    \qquad
    \subfloat[]{\includegraphics[width=0.45\textwidth]{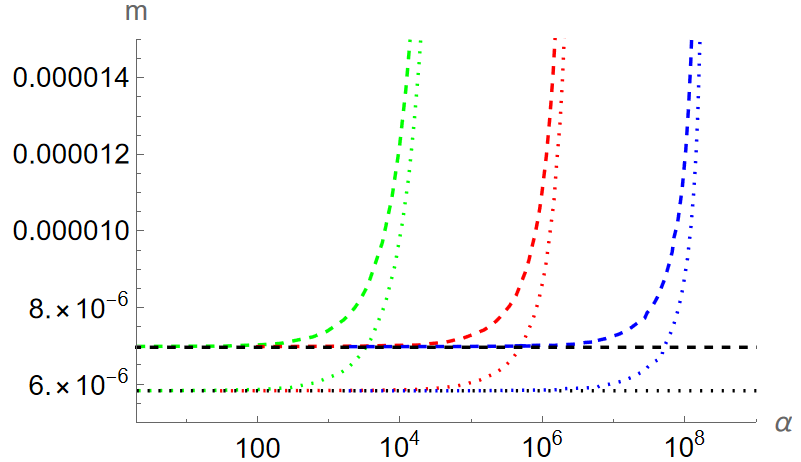}}%

    \caption{The observables $r$ vs. $n_s$ (a), $r$ vs. $\alpha$ (b), $\alpha$ vs. $n_s$ (c) and $m$ vs. $\alpha$ (d) for the model in \eqref{eq:testscenario} with $n=3/2$ (green),  $n= 7/4$ (red), $n=31/16$ (blue) and $n=2$ (black) for $N_e=50$ (dashed) and $N_e=60$ (dotted). The scalar amplitude $A_s$ is fixed to its observed value. For reference, we also plot the predictions for $N_e \in [50,60]$ of the corresponding asymptotic solutions in eqs. \eqref{eq:r:limit}--\eqref{eq:As:limit} (same color code, continuous line) and of quadratic inflation (black). The orange areas represent the 1,2$\sigma$ allowed regions coming  from  the latest combination of Planck, BICEP/Keck and BAO data \cite{BICEP:2021xfz}.}
    \label{fig.R+R^n}
\end{figure}

 In this subsection we apply our method on the model in \eqref{eq:testscenario} 
for $1 < n < 2$. Such a scenario satisfies all the requirements listed in \eqref{eq:low:energy:limit}--\eqref{eq:high:energy:limit}.
Applying \eqref{eq:r_zeta}, \eqref{eq:ns_zeta} and \eqref{eq:As_zeta} we can compute the phenomenological parameters:
\begin{eqnarray}
 N_e &=& \, \left[ \frac{\zeta  \left(n-(n-1) \right)}{8 m^2} \, _2F_1\left(1,\frac{1}{n-1};\frac{n}{n-1};(n-2) \alpha  \zeta ^{n-1}\right) \right]^{\zeta=\zeta_N}_{\zeta=\zeta_f} \, , \label{eq:Ne:zeta} \\
 r(\zeta_N) &=& \frac{64 m^2}{\zeta_N} \frac{1+\alpha (2-n)\zeta_N^{n-1}}{1+\alpha n \zeta_N^{n-1}} \,  , \label{eq:r:zeta:N}
 \end{eqnarray}
\begin{eqnarray}
n_s(\zeta_N) &=& 1 - \frac{8m^2}{\zeta_N}\frac{2+\alpha(n-2)(n-3)\zeta_N^{n-1}}{1+\alpha (2-n) n \zeta_N^{n-1}} \, , \label{eq:ns:zeta:N} \\
A_s(\zeta_N) &=&  \frac{1}{384 \pi ^2 m^2}
\frac{\zeta_N^2}{1+\alpha (2-n) \zeta _N^{n-1}} \label{eq:As:zeta:N} \, ,
\end{eqnarray}
where we used the hypergeometric function 
\be
_2F_1(a,b,c,z)=\sum_{k=0}^\infty \frac{(a)_n (b)_n}{(c)_n} \frac{z^n}{n!} \label{eq:2F1}
\ee
with $(q)_n$  the (rising) Pochhammer symbol. Notice that the $n=2$ case is also described by eqs. \eqref{eq:Ne:zeta}--\eqref{eq:As:zeta:N}.
We can also derive more readable expressions in the limit $|n-2| \alpha \rightarrow \infty$
(which automatically excludes the $n=2$ configuration). In such a limit we can approximate the number of $e$-folds as
\be
N_e \sim \frac{n}{8m^2} \zeta_N \, ,
\ee
obtaining
\begin{eqnarray}
 r(\zeta_N) &\simeq& \frac{8 (2-n)}{N_e} \, \label{eq:r:limit},\\
n_s(\zeta_N) &\simeq& 1 - \frac{3-n}{N_e} \, \label{eq:ns:limit} ,\\
A_s(\zeta_N) &\simeq& \frac{2^{2-3 n} \left(\frac{n}{N_e}\right)^{n-3}}{3 \pi^2   (2-n)} 
\frac{\left(m^2\right)^{2-n}}{\alpha} \label{eq:As:limit}
\, .
\end{eqnarray}
We show in Fig. \ref{fig.R+R^n} a more detailed numerical analysis, where we plot $r$ vs. $n_s$ (a), $r$ vs. $\alpha$ (b), $\alpha$ vs. $n_s$ (c) and $m$ vs. $\alpha$ (d) for $n=3/2$ (green),  $n= 7/4$ (red), $n=31/16$ (blue) with $N_e=50$ (dashed) and $N_e=60$ (dotted), while $A_s$ is fixed to the observed value \eqref{eq:As:constraint}.  For reference we also plot the predictions for $N_e \in [50,60]$ of the corresponding asymptotic solutions in eqs. \eqref{eq:r:limit}--\eqref{eq:As:limit} (same color code, continuous line), of quadratic inflation (black) and for $n=2$  with $N_e=50$ (black, dashed) and $N_e=60$ (black, dotted). The orange areas represent the 1,2$\sigma$ allowed regions coming  from  the latest combination of Planck, BICEP/Keck and BAO data \cite{BICEP:2021xfz}. The numerical results in Fig. \ref{fig.R+R^n}(a) were obtained by varying the parameter $m$ from $m = 6.98\cdot 10^{-6}$ ($N_e=50$) and $m = 5.82\cdot 10^{-6}$ ($N_e=60$) to $m = 1.6\cdot 10^{-5}$ (both $N_e=50,60$). This is equivalent to increase the parameter $\alpha$ since the relation between the two parameters is fixed by the amplitude of the power spectrum (cf. eqs. \eqref{eq:As_zeta} and \eqref{eq:As:limit}), as shown in Fig. \ref{fig.R+R^n}(d). From Fig. \ref{fig.R+R^n}(b) we notice that the net effect of the $\alpha R^n$ term is to lower the tensor-to-scalar ratio $r$.
As we get closer to $n=2$ we see that this effect is enhanced, as expected,
approaching the asymptotic value \eqref{eq:r:limit} for $\alpha \rightarrow \infty$. A similar discussion also applies to $n_s$ in Fig. \ref{fig.R+R^n}(c). In this case, $n_s$ increases until $\alpha$ is big enough, and then $n_s$ approaches the asymptotic value \eqref{eq:ns:limit}.

To conclude, we stress that the limits for $\alpha \to +\infty$ with $n=2$ and $n \neq 2$ are two completely different configurations. However, as can be deduced from Fig. \ref{fig.R+R^n}, keeping $n$ fixed, it is always possible to identify a maximum value for $\alpha$ so that the results between $n=2$ and $n \neq 2$ are indistinguishable and this maximum value increases with $n$ getting closer to 2. Then, for $\alpha$ values above such a maximum, the results for $n=2$ and $n \neq 2$ depart, converging towards two different asymptotic configurations. In particular, eqs. \eqref{eq:r:zeta:N} and \eqref{eq:ns:zeta:N} with $n=2$ are asymptotic limits, approached when $n$ is very close to 2, but never reached. Our results show that at a given $\alpha \gg 1$, a slight variation from $n=2$ might completely jeopardize the inflationary predictions of the $n=2$ case. This happens when $\alpha \gg \frac{1}{|n-2|}$.

\subsection{$n>2$} \label{sec:n_larger_than_2}

\begin{figure}[t]%
    \centering
    \subfloat[]{\includegraphics[width=0.45\textwidth]{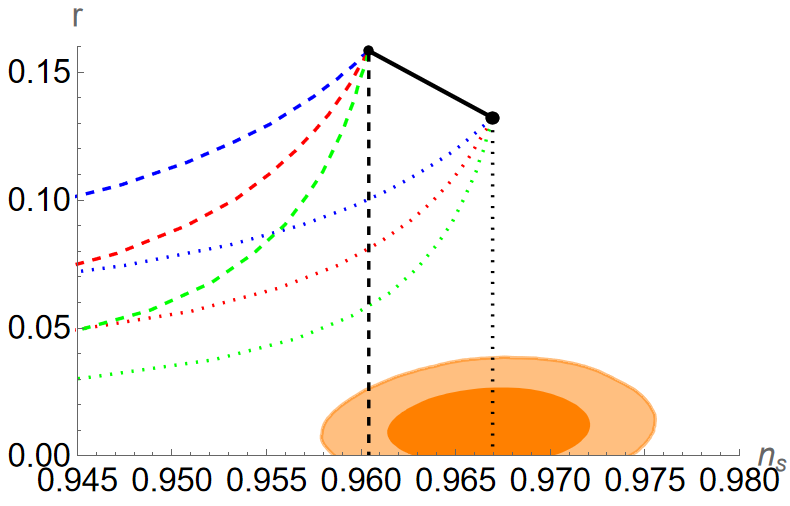}}%
    \qquad
    \subfloat[]{\includegraphics[width=0.45\textwidth]{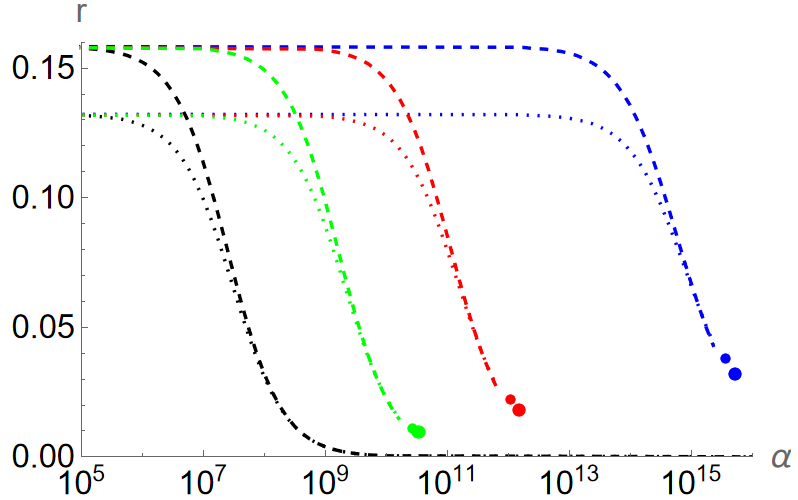}}%
    
    \subfloat[]{\includegraphics[width=0.45\textwidth]{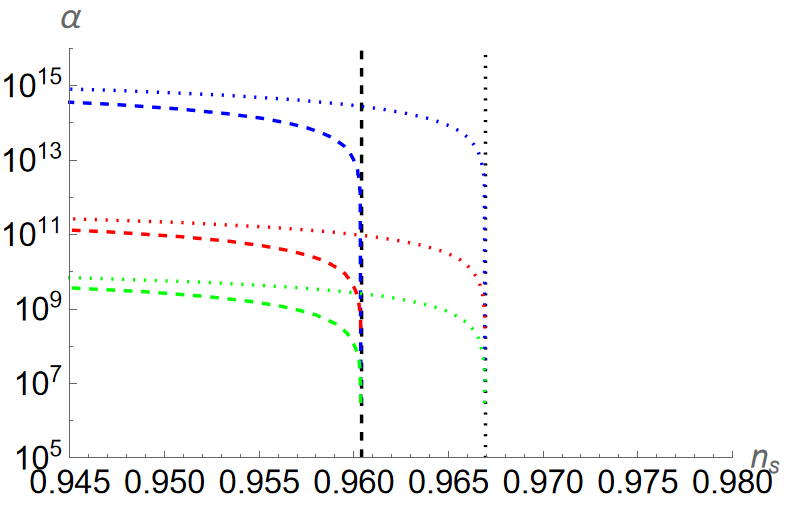}}%
    \qquad
    \subfloat[]{\includegraphics[width=0.45\textwidth]{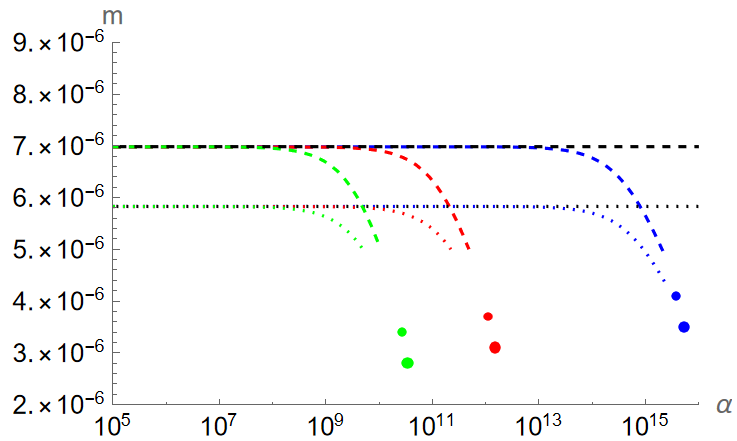}}
    \caption{The observables $r$ vs. $n_s$ (a), $r$ vs. $\alpha$ (b), $\alpha$ vs. $n_s$ (c) and $m$ vs. $\alpha$ (d) for the model in \eqref{eq:testscenario} with $n=3$ (blue),  $n= 5/2$ (red), $n=9/4$ (green) and $n=2$ (black) for $N_e=50$ (dashed) and $N_e=60$ (dotted). In the same color code, we show the limit values $r_{\bar\alpha}$ (eq. \eqref{eq:r:zetamax}) and $m_{\bar\alpha}$ (eq. \eqref{eq:m2:zetamax}) for $N_e=50$ (small dot) and $N_e=60$ (large dot). The scalar amplitude $A_s$ is fixed to its observed value.
    The orange areas represent the 1,2$\sigma$ allowed regions coming  from  the latest combination of Planck, BICEP/Keck and BAO data \cite{BICEP:2021xfz}.}
     \label{fig.n>2}
\end{figure}

 In this subsection, we test our method with the setup of eq. \eqref{eq:testscenario} but for $n > 2$. Such a scenario satisfies the requirements \eqref{eq:low:energy:limit}--\eqref{eq:Vinvertible}, but not \eqref{eq:high:energy:limit}. Moreover also \eqref{eq:Gbijective} requires additional care. We start by checking the constraint \eqref{eq:EoMzeta}:
\be
G(\zeta)=\frac{1}{4} \left[\zeta -\alpha  (n-2) \zeta ^n\right] = V(\phi) \, .
\ee
For $n>2$, $G(\zeta)$ is a function unbounded from below and exhibits a local maximum at $\zeta_\text{max}=[\alpha \, n \, (n-2) ]^{\frac{1}{1-n}}$. Therefore \eqref{eq:high:energy:limit} is not satisfied, but we can still treat the model as an effective theory if slow-roll is realized between 0 and $\zeta_\text{max}$. In such a domain we also manage to satisfy \eqref{eq:Gbijective}.

In the region of validity of the model the inflationary parameters are still given by \eqref{eq:Ne:zeta}--\eqref{eq:As:zeta:N}. We show in Fig. \ref{fig.n>2} a more detailed numerical analysis, where we plot $r$~vs.~$n_s$~(a), $r$~vs.~$\alpha$~(b), $\alpha$~vs.~$n_s$~(c) and $m$~vs.~$\alpha$~(d) for $n=3$ (blue),  $n= 5/2$ (red), $n=9/4$ (green) with $N_e=50$ (dashed) and $N_e=60$ (dotted). For reference we also plot the results for $n=2$  with $N_e=50$ (black, dashed) and $N_e=60$ (black, dotted). The scalar amplitude $A_s$ is fixed to its observed value. The orange areas again represent the 1,2$\sigma$ allowed regions coming  from  the latest combination of Planck, BICEP/Keck and BAO data \cite{BICEP:2021xfz}. The numerical results were obtained by varying the parameter $m$ in the range $ 3.95\cdot 10^{-6} < m < 5.82\cdot 10^{-6} $ ($N_e=60$) and $4.44\cdot 10^{-6} < m < 6.98\cdot 10^{-6} $ ($N_e=50$). Once again the net effect of the $\alpha R^n$ term is to lower $r$, and this effect is more enhanced as $n$ approaches 2. However, the effect on $n_s$ and $m$ is the opposite with the respect to the $n<2$ case. In fact, now, by increasing $\alpha$ both $n_s$ and $m$ are decreasing.

We can also see that, for a given $n$, there is an upper limit on $\alpha$. Since slow-roll inflation happens between  0 and $\zeta_\text{max}$, the possible number of e-folds is bounded from above in a given model. In order to get the required amount of $e$-folds, we need $\zeta_N < \zetamax$ at  $N_e \in [50,60]$. However, by increasing $\alpha$, $\zetamax$ decreases and the distance between $\zeta_N$ and $\zetamax$ decreases as well. We can set a rough upper limit for $\alpha$ when $\zeta_N = \zetamax$. The limit is only rough because $\eta$ has a pole at $\zeta=\zetamax$ meaning the loss of validity of the slow-roll approximation. Such a pole is reflected in Fig. \ref{fig.n>2}(c) with the appearance of horizontal asymptotes with $n_s$ pointing towards $-\infty$. Therefore the actual upper limit $\bar\alpha$ takes place for $\zeta_N$ not equal, but slightly smaller than $\zetamax$. Nevertheless, we can still provide useful estimates for the limit values of $r$, $m$ and $\alpha$ by using $\zeta_N = \zetamax$. First of all, we impose such a condition on the amplitude of the power spectrum \eqref{eq:As:zeta:N}, obtaining a limit for $m^2$:
\be
 m^2_{\bar\alpha} \simeq \frac{n (\bar\alpha  (n-2) n)^{-\frac{2}{n-1}}}{384 \pi ^2 (n-1) A_s} \label{eq:m2:zetamax} \, ,
\ee
where $A_s$ satisfies \eqref{eq:As:constraint}. We can now compute the number of $e$-folds \eqref{eq:Ne:zeta} till $\zetamax$, obtaining
\be
 N_e \sim (\bar\alpha  (n-2)
   n)^{\frac{1}{n-1}} \, \frac{48 \pi ^2 (n-1) A_s}{n} \, \left[n+(1-n) \, _2F_1\left(1,\frac{1}{n-1};\frac{n}{n-1};\frac{1}{n}\right)\right] \, . \label{eq:Ne:zetamax}
\ee
We can use \eqref{eq:Ne:zetamax} as a definition for $\bar\alpha$. Using the previous results, we obtain the limit value for $r$ 
\be
 r_{\bar\alpha} \simeq \frac{(n-2) (\bar\alpha  (n-2) n)^{\frac{1}{1-n}}}{6 \pi ^2 (n-1) A_s} \, . \label{eq:r:zetamax}
\ee
The limit values $r_{\bar\alpha}$ and $m_{\bar\alpha}$ are shown respectively in Fig. \ref{fig.n>2}(b) and Fig. \ref{fig.n>2}(d) for $n=3$ (blue), $n=5/2$ (red) and $n=9/4$ (green). The small (large) dot stands for $N_e=50$ (60) $e$-folds. As we can see, the numerical values for $r$ approach closely the limit ones, but cannot reach them because that would imply a violation of  the slow-roll approximation. Analogously,  $m_{\bar\alpha}$ and the actual limit of $m$ have the same order of magnitude.

\section{Beyond slow-roll approximation}
\label{sec:beyond_SR}
To gain a sense of the global dynamics of our models, it is interesting to solve their evolution numerically without the slow-roll approximation, in particular for the problematic $n>2$ case. Starting from the action \eqref{eq:action:zeta:E}, after some manipulations, the full Einstein frame EoMs read:
\begin{gather}
    \label{eq:phi_eom}
    \ddot{\phi} + 3H\dot{\phi} + \frac{V'(\phi)}{F'(\zeta)k(\phi)} = \frac{\dot{\phi}\dot{\zeta}F''(\zeta)}{F'(\zeta)} - \frac{1}{2}\frac{k'(\phi)}{k(\phi)}\dot{\phi}^2 \, , \\
    \label{eq:H_eom}
    3H^2 = \frac{1}{2}\frac{\dot{\phi}^2}{F'(\zeta)}k(\phi) + U(\phi,\zeta) \, , \\
    \label{eq:zeta_eom}
    -\frac{1}{2}\dot{\phi}^2 F'(\zeta)k(\phi) + 2V(\phi) - 2G(\zeta) = 0 \, .
\end{gather}
These can be used to also derive
\begin{gather}
    \dot{H} = -\frac{1}{2}\frac{\dot{\phi}^2}{F'(\zeta)}k(\phi) \, , \\
    \epsilon_H \equiv -\frac{\dot{H}}{H^2} = \frac{12V(\phi) - 6F(\zeta) + 3\zeta F'(\zeta)}{6V(\phi) - 3F(\zeta) + 2\zeta F'(\zeta)} \, .
\end{gather}
Again, solving the constraint equation \eqref{eq:zeta_eom} may be problematic, but it can be replaced with its time derivative which, using \eqref{eq:phi_eom} and \eqref{eq:zeta_eom}, reads
\begin{equation} \label{eq:zeta_deriv}
    \dot{\zeta} = \frac{3H\dot{\phi}^2 F'(\zeta) k(\phi) + 3 V'(\phi)\dot{\phi}}{2G'(\zeta) + \frac{3}{2}\dot{\phi}^2 F''(\zeta)k(\phi)} \, .
\end{equation}
To solve the full time evolution, one only needs to solve \eqref{eq:zeta_eom} once to get the initial condition of $\zeta$; after that, it is simple to follow the time evolution of $\zeta$ through \eqref{eq:zeta_deriv}. The constraint \eqref{eq:zeta_eom} can later be used to check the accuracy of the result.

Let us check the slow-roll limit of the full equations. There, the potential terms dominate over the kinetic ones. The Hubble constraint becomes $3H^2=U$ as usual. The constraint \eqref{eq:zeta_eom} becomes \eqref{eq:EoMzeta}, $G(\zeta)=V(\phi)$, fixing a one-to-one correspondence between $\phi$ and $\zeta$ and the new canonical field $\chi$, which in this limit can be defined through \eqref{eq:dchidphi}. Since in this limit, $\pd_\zeta U(\phi,\zeta) = 0$ (by construction), we have 
\begin{equation}
    \frac{\dd U}{\dd \chi} = \frac{\dd \phi}{\dd \chi}\partial_\phi U(\phi,\zeta) = \frac{V'(\phi)}{F'(\zeta)^{3/2}k(\phi)^{1/2}} \, ,
\end{equation}
and the field equation becomes
\begin{equation}
    3H\dot{\phi} = -\frac{V'(\phi)}{F'(\zeta)k(\phi)} \qquad \Rightarrow \qquad 3H\dot{\chi} = -\frac{\dd U}{\dd \chi} \, ,
\end{equation}
as expected. In practice, the goodness of the slow-roll approximation can be estimated by comparing the extra terms in \eqref{eq:phi_eom}, \eqref{eq:H_eom}, and \eqref{eq:zeta_eom} to the leading slow-roll terms.

\subsection{$n<2$}

\begin{figure}[t]
    \centering
    \includegraphics[scale=0.9]{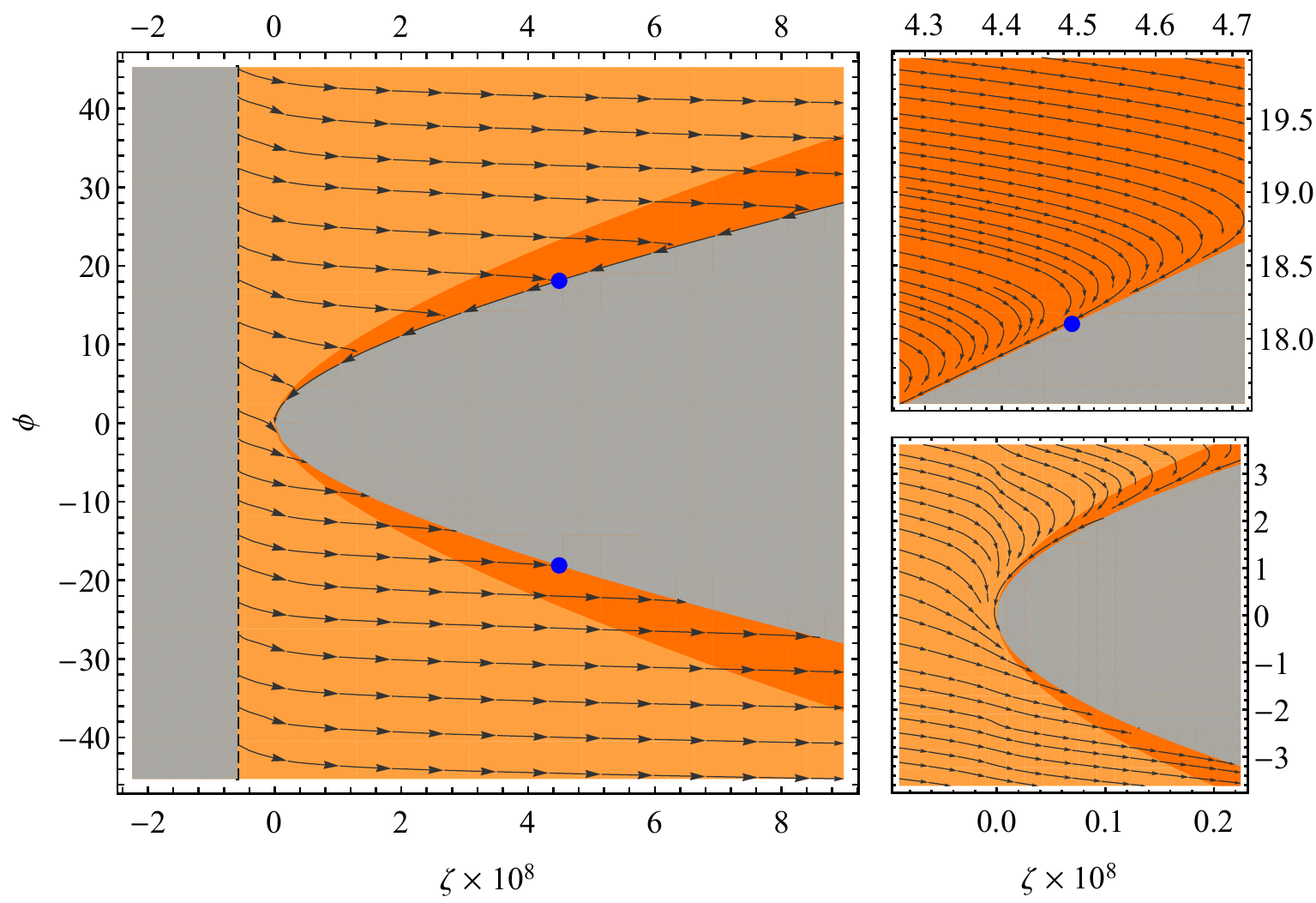}
    \caption{Time evolution of the model \eqref{eq:testscenario} for the benchmark point in \eqref{eq:benchmark:32} in the $(\zeta,\phi)$ plane as given by \eqref{eq:zeta_eom}, \eqref{eq:zeta_deriv} (the $\dot{\phi}<0$, $H>0$ branch).  The grey areas are excluded because either $F'(\zeta)<0$ or \eqref{eq:zeta_eom} can't be satisfied. The darker orange region corresponds to inflation. The blue dots denote the CMB scale with $N_e=50$, and have $A_s = 2.1 \cdot 10^{-9}$, $n_s=0.967$, $r=0.096$. 
    }
    \label{fig:flow_3/2}
\end{figure}

In this subsection, we study the time evolution beyond the slow-roll approximation for the test scenario in \eqref{eq:testscenario} with $n<2$. In particular, we consider the following benchmark point\footnote{To deal with the fractional exponent for negative $\zeta$, we take $F(R) = R + \alpha|R|^{3/2}$.}${}^{,}$\footnote{We remind the reader that all parameters and field values are in Planck units, $\MP=1$.}:
\be
 n=3/2 \, , \quad \alpha=8710 \, , \quad m=1.15 \cdot 10^{-5} \, . \label{eq:benchmark:32}
\ee
The corresponding time evolution given by \eqref{eq:zeta_eom}, \eqref{eq:zeta_deriv} in the $(\zeta,\phi)$ plane is depicted in the flow chart of Fig. \ref{fig:flow_3/2}. The darker orange region corresponds to inflation with $\epsilon_H < 1$. As expected, inflation takes place only when $\zeta>0$ (cf. eq. \eqref{eq:Uzeta}). The grey areas represent excluded regions either because $F'(\zeta)$ turns negative and $\dot{\phi}$ diverges\footnote{At the same limit, the Weyl transformation \eqref{eq:Weyl} becomes singular.} (for $\zeta < -5.9\cdot 10^{-9}$), or because $V(\phi) < G(\zeta)$ and the constraint equation \eqref{eq:zeta_eom} does not have real solutions for $\dot{\phi}$ (for small $\phi$, large $\zeta$). Slow-roll happens at the edge of this region, where $V(\phi) \approx G(\zeta)$, as explained above. The blue dot denotes the CMB scale with $N_e=50$, and has $A_s = 2.1\cdot 10^{-9}$, $n_s=0.967$, $r=0.096$.

In Fig. \ref{fig:flow_3/2}, it was assumed that $H>0$ and $\dot{\phi} < 0$. This is only one branch of the possible solutions. However, from the EoMs and symmetry of the potential $V$ we see that the system stays invariant under the transformations $\dot{\phi} \leftrightarrow -\dot{\phi}$, $\phi \leftrightarrow -\phi$ (in particular, $\dot{\zeta}$ does not change). Thus, the $\dot{\phi}>0$ branch is obtained by mirroring Fig. \ref{fig:flow_3/2} with respect to the $x$-axis. The system can only jump from one branch to the other when $\dot{\phi}=0$, that is, at the slow-roll edge of the right hand side grey region. In Fig. \ref{fig:flow_3/2}, trajectories with $\phi<0$ end up on the lower edge and switch to the $\dot{\phi}>0$ branch: the field rolls up the potential, slows down, stops, and turns around, entering slow-roll on the other branch. Trajectories with $\phi>0$ approach the upper edge, but slow down due to Hubble friction and enter slow-roll right next to the edge on the same branch. This can be seen as sharp turns in the $(\zeta,\phi)$ trajectories in the top right panel of Fig. \ref{fig:flow_3/2}. Note that two more branches, with $H<0$, can be obtained by simply switching the direction of the flow; these can't be reached smoothly from the inflating branches in a spatially flat universe filled only with a scalar field.

Near the end of a slow-roll trajectory, the field approaches the origin with $\phi=\zeta=0$, oscillating around it. This can be seen in the bottom right panel of Fig. \ref{fig:flow_3/2}. Here $\dot{\phi}$ changes sign repeatedly, and the evolution jumps from one branch to another as the oscillation amplitude dies down due to Hubble friction. Note that $\zeta<0$ repeatedly during the oscillations.

Due to a rescaling symmetry of the action, the stream lines of a figure like \ref{fig:flow_3/2} exactly describe the dynamics of a family of models where $n$ and $\alpha \cdot m^{2(n-1)}$ are kept constant \cite{Tenkanen:2020cvw}, up to a linear rescaling of $\zeta$. This rescaling preserves the values of $N_e$, $n_s$, and $r$, but changes $A_s$. Beyond that, a chart that is qualitatively similar to Fig. \ref{fig:flow_3/2} can be drawn for any model of the form \eqref{eq:testscenario} with $1<n<2$. Remarkably, in these models all trajectories with large enough initial field values eventually end up in the inflationary region and on a slow-roll trajectory.

\subsection{$n>2$} \label{sec:n_larger_than_2_beyond_SR}

\begin{figure}[t]
    \centering
    \includegraphics[scale=0.9]{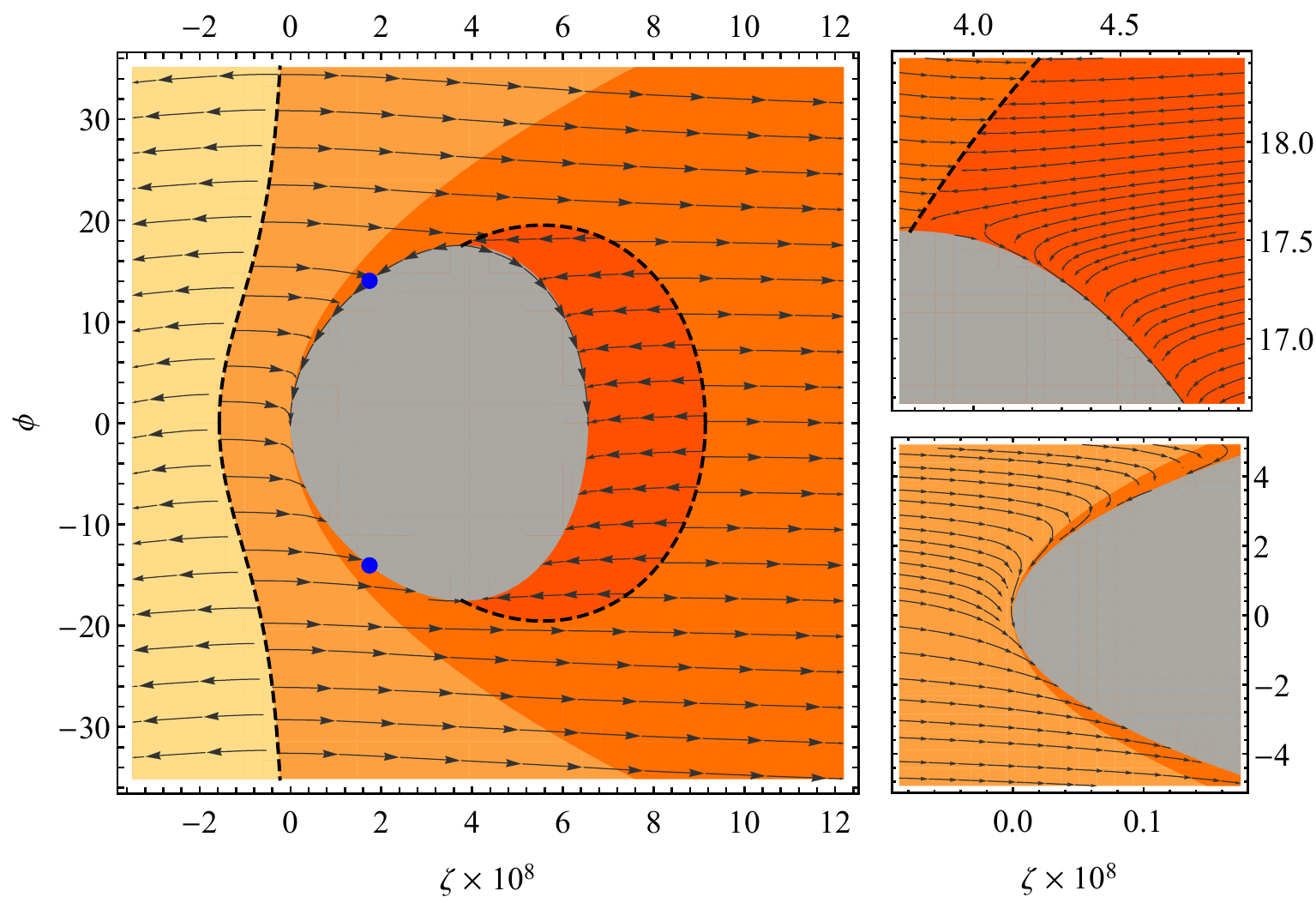}
    \caption{Time evolution of the model \eqref{eq:testscenario} for the benchmark point in \eqref{eq:benchmark:3} in the $(\zeta,\phi)$ plane as given by \eqref{eq:zeta_eom}, \eqref{eq:zeta_deriv} (the $\dot{\phi}<0$, $H>0$ branch). Same as Fig. \ref{fig:flow_3/2} with differences discussed in the text. The CMB numbers at the blue dot are $N_e=50$, $A_s = 2.1 \cdot 10^{-9}$, $n_s=0.952$, $r=0.116$.}
    \label{fig:flow_3}
\end{figure}

In this subsection we study the time evolution beyond the slow-roll approximation for the test scenario in \eqref{eq:testscenario} with $n>2$. In particular, we consider the following benchmark point:
\be
 n=3 \, , \quad \alpha=2.32\cdot10^{14} \, , \quad m=6.40\cdot10^{-6} \, . \label{eq:benchmark:3}
\ee
The corresponding time evolution given by \eqref{eq:zeta_eom}, \eqref{eq:zeta_deriv} (the $\dot{\phi}<0$, $H>0$ branch) in the $(\zeta,\phi)$ plane is depicted in Fig. \ref{fig:flow_3}. This is analogous to Fig. \ref{fig:flow_3/2}, but with the crucial difference that now $G(\zeta)$ has a maximum (here at $\zeta = 3.79 \cdot 10^{-8}$, corresponding to $|\phi|=17.5$ on the slow-roll line) and begins to decrease for larger $\zeta$, making the grey region where constraint \eqref{eq:zeta_eom} can't be satisfied terminate at $\zeta=6.62 \cdot 10^{-8}$. Note also that for $|\phi| > 17.5$, not all $\dot{\phi}$-values are allowed, but there's a minimum value of $|\dot{\phi}|$ needed to solve \eqref{eq:zeta_eom}. This limits the possible initial conditions of the system in terms of the Jordan frame variables.

A new feature emerges for $\zeta > 3.79 \cdot 10^{-8}$, marked by the dashed black line: a divergence of $\dot{\zeta}$ from \eqref{eq:zeta_deriv}, possible here since $G'(\zeta)<0$. The darkest orange region inside this curve is separated from the rest of phase space---as Fig. \ref{fig:flow_3} shows, it either attracts or repels all trajectories around it and can't be dynamically crossed---and while the system inflates there, and is even attracted to a slow-roll trajectory near the grey boundary as shown by the top right panel of Fig. \ref{fig:flow_3}, it is driven towards the point $(6.62 \cdot 10^{-8},0)$ where condition \eqref{eq:low:energy:limit} is broken and we don't have the usual Einstein-Hilbert low-energy limit for gravity.

We also have $G(\zeta)<0$ for $\zeta < 0$. The low $\zeta$ limit is not given by the condition $F'(\zeta)=0$ as above (here $F'(\zeta)>0$ everywhere), but instead another divergence of $\dot{\zeta}$ (this time caused by $F''(\zeta) < 0$), denoted by the left-most dashed line. We cut off Fig.~\ref{fig:flow_3} near this line since, once again, it can't be crossed dynamically, so the left hand side is cut off from the observationally allowed trajectories in the vicinity of $\phi=\zeta=0$.

These features are generic for models with $n>2$ and signify their unhealthiness. Slow-roll can still happen at the edge of the grey area and terminate succesfully near $\phi=\zeta=0$, as depicted by the bottom right panel of Fig. \ref{fig:flow_3}, though the maximum number of slow-roll e-folds is limited, as discussed above. However, most trajectories pass the slow-roll region and continue to $\zeta \to \infty$, never turning back. In this sense, the slow-roll region is not an attractor of the dynamics on a global scale.

\section{Conclusions} \label{sec:Conclusions}
We studied the dynamics of a (minimally coupled) single field inflaton in the presence of Palatini $F(R)$ gravity. Since such a scenario is not always explicitly solvable, we developed a method that allows the computation of the inflationary parameters if certain conditions are satisfied. Apart from the usual requirements of a generic $F(R)$ theory, as reproducing GR in the low energy limit or having attractive gravity, we found one additional constraint to be important: for curvatures going to infinity, either $F(R)$ should not diverge or it should not diverge faster than $R^2$.
In case this last requirement is not satisfied, the theory exhibits problematic UV behaviour with additional divergences in phase space, though it can treated as an effective theory during slow-roll inflation. Moreover, to successfully apply our procedure, the Jordan frame inflaton potential $V(\phi)$ has to be an invertible function of $\phi$. 

We applied our method to a test scenario of an inflaton with a canonical kinetic term and a quadratic potential embedded in $F(R)=R+\alpha R^n$ gravity. We computed the inflationary predictions for both the $n<2$ case, which satisfies all the requirements, and the $n>2$ case with problematic behaviour. Both cases share the same effect on the tensor-to-scalar ratio $r$: for $\alpha$ increasing, $r$ decreases, with the effect getting enhanced with $n$ approaching 2. For the other phenomenological parameters, the two cases have opposite behaviours: for $\alpha$ increasing when $n<2$ $(n>2)$, $n_s$ and $m$ are increasing (decreasing). Moreover, for $n>2$ and given $n$ and $N_e$, $\alpha$ shows an upper limit. 
We also checked the evolution of the system beyond the slow-roll approximation for both models. We discovered that $n<2$ behaves well: all trajectories with large enough initial field values eventually end up in the inflationary region and on a slow-roll trajectory. On the other hand, in the $n>2$ case, the slow-roll region is not an attractor of the dynamics on a global scale, another sign of the intrinsic illness of this setup.

We conclude with a remark about the $n=2$ scenario. Such a configuration has been quite a powerful tool to adjust inflationary models, reducing $r$ while leaving $n_s$ practically unchanged. However, our study proves that in the strong coupling limit $\alpha \gg 1$, a slight variation from $n=2$ can induce a large change in the $n_s$ predictions. This might have a dramatic impact in ruling in/out inflationary models, especially in light of the increased precision of future  experiments (e.g. Simons Observatory \cite{SimonsObservatory:2018koc}, PICO \cite{NASAPICO:2019thw}, CMB-S4 \cite{Abazajian:2019eic} and LITEBIRD \cite{LiteBIRD:2020khw}).


\acknowledgments

This work was supported by the Estonian Research Council grants MOBTT5, MOBTT86, PRG1055 and by the ERDF Centre of Excellence project TK133.

\appendix

\section{Comparison to S. Bekov et al.} \label{sec:comparison_to_other_RN_computation}
The authors of S. Bekov et al. \cite{Bekov:2020dww} also studied slow-roll inflation in Palatini $F(R)$ models, with the same potential $V(\phi)=\frac{1}{2}m^2\phi^2$ as our \eqref{eq:testscenario}, but choosing $F(R)=\alpha R^n$ without the linear part. As noted in section~\ref{subsec:requirements}, this may be problematic for not producing the right low-$R$ limit. Nevertheless, our results from section \ref{sec:test} should coincide with theirs in the limit $\alpha \to \infty$, but this is not the case. In particular, their prediction of $r=0.34$ for $n=3$ is significantly larger than ours, see Fig.~\ref{fig.n>2}.

One culprit for the difference is a sign error in equation~(49) of \cite{Bekov:2020dww}:
\begin{equation} \label{eq:Bekov_extra_minus}
    \phi(\chi) \approx \qty({\color{red}-}\frac{k(n-2)}{2m_\chi^2(n-1)}\frac{1}{\kappa^2\chi^2})^{\frac{1-n}{n}} \, .
\end{equation}
Solving for $\phi(\chi)$ from their equations (37) and (48) produces the extra minus sign, here in red, that they omitted. Note that we use a different notation, and a somewhat different formalism, from that of \cite{Bekov:2020dww}. Table \ref{tab:dictionary} presents a dictionary between the two computations.

\begin{table}
    \centering
    \begin{tabular}{cc}
         Our notation & Notation of S. Bekov et al. \cite{Bekov:2020dww} \\
         \hline
         $\phi$ & $\chi$ \\
         $V(\phi)=\frac{1}{2}m^2\phi^2$ & $U(\chi)=\frac{1}{2}m_\chi^2\chi^2$\\
         $R=\zeta$ & $\mathcal{R}$\\
         $F(\zeta)$ & $f(\mathcal{R})$\\
         $F'(\zeta)$ & $\phi$ \\
         $\zeta F'(\zeta) - F(\zeta)$ & $V(\phi)=\mathcal{R} \phi - f(\mathcal{R})$ \\
         $\zeta$ & $V'(\phi)$ \\
         $G(\zeta)=\frac{1}{4}\qty[2F(\zeta)-\zeta F'(\zeta)]$ & $-\frac{1}{4}\qty[2V(\phi)-\phi V'(\phi)]$ \\
         $\sqrt{F'(\zeta)}\dot{\phi}$ & $\dot{\chi}$
    \end{tabular}
    \caption{A dictionary between the notation of this article and that of \cite{Bekov:2020dww}. Note, in particular, the relation of the time derivatives: we use the Einstein frame time, whereas \cite{Bekov:2020dww} works in the Jordan frame, leading to a difference related to the conformal factor $F'(\zeta)$. We have set $k(\phi)=1$, and worked in Planck units, $M_P \equiv \kappa^{-1}=1$.}
    \label{tab:dictionary}
\end{table}

The minus sign in \eqref{eq:Bekov_extra_minus} makes $\phi(\chi)$ complex for $n>2$, providing no slow-roll solutions. Indeed, for $n>2$ their constraint equation for the auxiliary field in (35),
\begin{equation} \label{eq:Bekov_constraint}
    2V(\phi)-\phi V'(\phi) = \kappa^2\qty(\dot{\chi}^2 - 4U(\chi)) \, ,
\end{equation}
can't be solved in the slow-roll limit of small field velocities. Using the dictionary of table~\ref{tab:dictionary} we see that \eqref{eq:Bekov_constraint} corresponds to our constraint equation \eqref{eq:EoMzetafull}, or equivalently \eqref{eq:zeta_eom}, and for $n>2$, we have $G(\zeta)=\frac{\alpha}{4}(2-n)\zeta^n<0$ for all $\zeta>0$, explaining the inability to find a solution. We see similar behaviour in our $F(R) = R + \alpha R^n$ case, where $G$ becomes negative when the $\alpha R^n$ term starts to dominate at large field values, see the discussion in sections \ref{subsec:requirements}, \ref{sec:n_larger_than_2}, and \ref{sec:n_larger_than_2_beyond_SR}. Indeed, in our slow-roll results in section~\ref{sec:n_larger_than_2}, slow-roll always happens in the small-field regime where the linear $R$ term is still significant.

In addition to the sign error, the authors of \cite{Bekov:2020dww} used slow-roll parameters computed in the Jordan frame in the standard CMB formulae \eqref{eq:r_zeta}, \eqref{eq:ns_zeta}. However, these formulae assume metric Einstein-Hilbert gravity with a canonical scalar field, and thus only work with slow-roll parameters computed in the Einstein frame as we did in our sutdy. These problems render the results of \cite{Bekov:2020dww} invalid and explain the differences between our results and theirs.

\bibliographystyle{JHEP}
\bibliography{references}

\end{document}